\newcommand{\MGEV}{\mbox{GeV/c$^{2}$}}
\newcommand{\EGEV}{\mbox{GeV/c}}
\newcommand{\PGEV}{\mbox{GeV$^2$/c$^{2}$}}

\newcommand{\be}	{\begin{equation}}
\newcommand{\ee}	{\end{equation}}
\newcommand{\pip}       {\mbox{$\pi^{+}$}}
\newcommand{\pim}       {\mbox{$\pi^{-}$}}

\newcommand{\et}       	{\mbox{$\eta\/$}}

\newcommand{\reaction}  {\mbox{$\pi^{-}p \rightarrow \eta \pi^{+} \pi^{-} n\/$}}  

\newlength{\captsize}		\let\captsize=\footnotesize
\newlength{\captwidth}		
\newlength{\beforetableskip}	

\newcommand{\capt}[1]{\begin{minipage}{\captwidth}
	\let\normalsize=\captsize
	\caption{#1}
	\end{minipage}\\ \vspace{\beforetableskip}}
\documentstyle[preprint,aps, epsfig]{revtex}
\begin{document}

\newcounter{TC}
\newcommand{\useTC}[1]{\refstepcounter{TC}\label{#1}\arabic{TC}}

\draft
\title{Partial-Wave Analysis of the $\eta \pi^+ \pi^-$
System Produced in the Reaction $ \pi^{-} p \rightarrow \eta \pi^+
\pi^- n$ at 18 \EGEV}
\author{
J.~J.~Manak,\footnote[1]{Present address: Thomas Jefferson
National Accelerator Facility, Newport News, VA 23606, USA}
T.~Adams, \footnote[2]{Present address: Department of Physics,
Kansas State University, Manhattan, KS 66506, USA} J.~M.~Bishop,
N.~M.~Cason, E.~I.~Ivanov, J.~M.~LoSecco,
 A.~H.~Sanjari, W.~D.~Shephard,
D.~L.~Stienike, S.~A.~Taegar,\footnote[3]{Present address:
Department of Physics, University of Arizona, Tucson, AZ 85721,
USA} D.~R.~Thompson }
\address{\it Department of Physics, University of Notre Dame,
Notre Dame, IN 46556, USA}
\author{S.~U.~Chung, K.~Danyo,
R.~W.~Hackenburg, C.~Olchanski, D.~P.~Weygand,\footnotemark[1]
H.~J.~Willutzki}
\address{\it Department of Physics, Brookhaven National Laboratory,
Upton, NY 11973, USA}
\author{
A.~R.~Dzierba,
J.~Gunter,
R.~Lindenbusch, D.~R.~Rust, E.~Scott, P.~T.~Smith, T.~Sulanke,
S.~Teige }
\address{\it Department of Physics, Indiana University,
Bloomington, IN 47405, USA}
\author{
S.~P.~Denisov, V.~A.~Dorofeev, I.~A.~Kachaev, V.~V.~Lipaev,
A.~V.~Popov,
 D.~I.~Ryabchikov
}
\address{\it Institute for High Energy Physics, Protvino, Russian Federation}
\author{
Z.~Bar-Yam, J.~P.~Dowd, P.~Eugenio,\footnote[4]{Present address:
Department of Physics, Carnegie Mellon University, Pittsburgh, PA
15213, USA} M.~Hayek,\footnote[5]{Permanent address:  Rafael,
Haifa, Israel} W.~Kern, E.~King, N.~Shenhav\footnotemark[5] }
\address{\it Department of Physics, University of Massachusetts Dartmouth,
North Dartmouth, MA 02747, USA}
\author{
V.~A.~Bodyagin, O.~L.~Kodolova, V.~L.~Korotkikh, M.~A.~Kostin,
A.~I.~Ostrovidov, L.~I.~Sarycheva, N.~B.~Sinev, I.~N.~Vardanyan,
A.~A.~Yershov }
\address{\it Institute for Nuclear Physics, Moscow State University,
Moscow, Russian Federation}
\author{
D.~S.~Brown,\footnote[6]{Present address: Department of Physics,
University of Maryland, College Park, MD 20742, USA} T.~K.~Pedlar,
K.~K.~Seth, J.~Wise, D.~Zhao }
\address{\it Department of Physics, Northwestern University,
Evanston, IL 60208, USA}
\author{
G.~S.~Adams, J.~P.~Cummings, J.~Kuhn, J.~Napolitano, M.~Nozar,
J.~A.~Smith, D.~B.~White, M.~Witkowski }
\address{\it Department of Physics, Rensselaer Polytechnic Institute.
Troy, NY 12180, USA}
\author{
(E852 Collaboration)}
\date{\today}
 \maketitle
\begin{abstract}
A partial-wave analysis of 9082 $\eta\pi^+\pi^- n $ events
produced in the reaction $\pi^{-} p \rightarrow \eta\pi^+\pi^- n$
at $18.3$ \EGEV\ has been carried out using data from experiment
852 at Brookhaven National Laboratory. The data are dominated by
$J^{PC} = 0^{-+}$ partial waves consistent with observation of the
$\eta(1295)$ and the $\eta(1440)$. The mass and width of the
$\eta(1295)$ were determined to be $1282 \pm 5$~MeV and $66 \pm
13$~MeV respectively while the $\eta(1440)$ was observed with a
mass of $1404 \pm 6$~MeV and a width of $80 \pm 21$~MeV. Other
partial waves of importance include the $1^{++}$ and the $1^{+-}$
waves. Results of the partial wave analysis are combined with
results of other experiments to estimate $f_1(1285)$\ branching
fractions.  These values are considerably different from current
values determined without the  aid of amplitude analyses.
\end{abstract}


\section{Introduction}

In this paper we present results of a partial-wave analysis of the
$\eta\pi^+\pi^-$ system in the 1210 to 1530~${\rm MeV}/c^2$ mass
region, as obtained from the study of  the reaction
\begin{equation}
\label{reaction}
\pi^{-} p \rightarrow \eta\pi^+ \pi^- n,~~\eta \rightarrow 2\gamma
\end{equation}
at 18.3~\EGEV. The data sample was collected during the summer of
1994 using the Multi-Particle Spectrometer (MPS) at the
Alternating Gradient Synchrotron (AGS) facility of Brookhaven
National Laboratory (BNL).

The identification of the isoscalar members of the $J^{PC} =
0^{-+}$ and $1^{++}$ nonets has been the subject of considerable
interest, particularly with regard to searches for exotic mesons.
It is known that such states often have $a_0(980)\pi$ decay modes.
Since the  $a_0(980)$ couples to both $\eta\pi$ and to
$K\overline{K}$ final states, comparison of the resonances
produced in the $\eta\pi^+\pi^-$ and $K \overline{K} \pi$\
reactions can lead to important information with regard to this
identification.

The \et\pip\pim\ system is complicated, characterized by the large
range of accessible quantum numbers ($J^{PC}=0^{-+}$, $0^{--}$,
$1^{--}$, $1^{+-}$, $1^{++}$, $2^{--}$ \dots), a large number of
possible $\eta \pi$\ and $\pi \pi$\ intermediate isobars
($\sigma$, $\rho(770)$, $a_{0}(980)$, $f_{2}(1270)$,
$a_{2}(1320$))\footnote{We refer to the $\pi\pi$ S-wave as
$\sigma$.  The form used for this is discussed in Section
~\ref{fitting}.}, and the presence of overlapping resonances
($f_{1}(1285)$\/, $\eta(1295)$).

Historically, the low-mass region around the 1300~${\rm MeV}/c^2$
enhancement in the $\eta\pi^+\pi^-$ and $K \overline{K} \pi$\ mass
spectra was called the D region. Most early analyses
~\cite{ch1:campbell,ch1:corden,ch1:Dahl,ch1:nacasch,ch1:defoix,ch1:gurtu}
 made the assumption that a single state existed in this region
in the presence of an incoherent (non-interfering) background. The
problem was then the determination of the appropriate quantum
numbers of this state and its branching ratio to \et \pip \pim.
Most early experiments showed a preference for $J^{PC} = 1^{++}$
quantum numbers for this state, now referred to as the
$f_1(1285)$~\cite{book:pdb}.

Later, sufficiently high statistics were collected to carry out a
partial wave analysis of the $\eta \pi^+ \pi^-$ system. Stanton
{\em et al.}~\cite{ch1:stanton} performed an analysis of the
reaction $\pi^{-} p \rightarrow \eta \pi^+ \pi^- n$ at 8.45~\EGEV.
The low-mass region was fit with a combination of $0^{-+}$,
$1^{++}$, and $1^{+-}$ partial waves. Their analysis suggested the
presence of a new state with $J^{PC} = 0^{-+}$, the $\eta(1295)$,
as well as the $f_1(1285)$. In addition, it was suggested that the
fit could be improved considerably if the $0^{-+}$\ partial waves
were not allowed to interfere with the other waves in the fit.

The KEK-E179 collaboration performed two partial wave
analyses~\cite{ch1:fukui,ch1:ando} of the same reaction at
8~\EGEV. They too used a set of $0^{-+}$, $1^{++}$, and $1^{+-}$
partial waves to describe the low-mass region, and observed the
$f_1(1285)$\ and $\eta(1295)$. Their analysis also suggested the
presence of an additional state, now called~\cite{book:pdb} the
$\eta(1440)$,
 in the high-mass region, which was earlier called the E region.

\section{Apparatus}

\begin{center}
\begin{figure}
    \psfig{figure=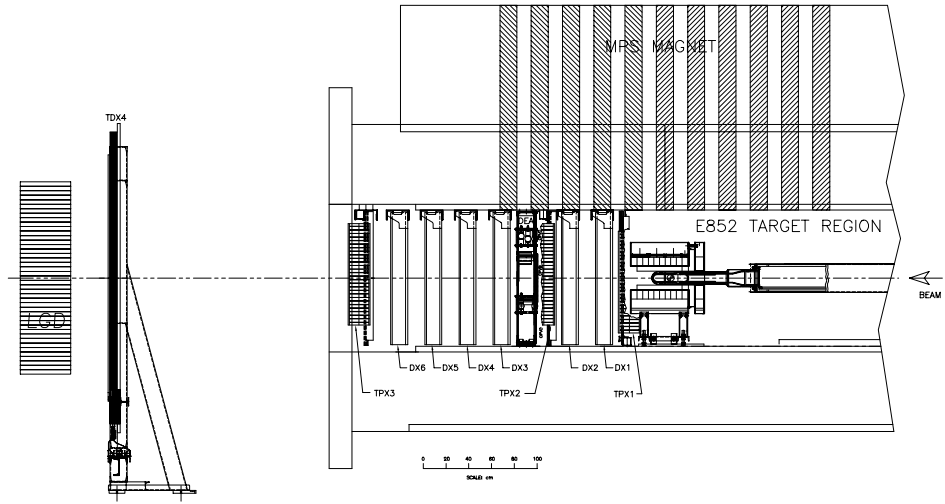,width=7.5cm,angle=90}
    \caption{E852 Apparatus Layout
\label{layout}}
\end{figure}
\end{center}

Figure~\ref{layout} shows the elevation view of the experimental
layout. The detector system consists of a charged-particle
spectrometer and a downstream 3045-element lead-glass
electromagnetic calorimeter
(LGD)~\cite{crittendon}~\cite{crittendon2} to provide neutral
particle detection.

An 18.3~\EGEV\ $\pi^-$ beam is incident on the 30-cm
liquid-hydrogen target located at the center of the MPS magnet.
Three threshold \^{C}erenkov counters in the beam line are used to
tag the beam particles as pions.   Surrounding the target is a
198-element thallium-doped cesium iodide cylindrical veto array
(CsI)~\cite{csi} used in off-line analysis to reject events with
wide-angle, low-energy photons from the decays of baryonic
resonances. Between the target and the CsI is a four-plane
cylindrical drift chamber (TCYL)~\cite{tcyl} for triggering on
recoil charged tracks. The downstream half of the magnet is
equipped with three proportional wire chambers (TPX1-3) for
triggering on forward charged-track multiplicity, and six drift
chamber modules (DX1-6) for measuring the momentum of forward
charged tracks. Also in the magnet is a window-frame
lead-scintillator sandwich photon veto counter (DEA) which covers
the solid-angle gap between the CsI and the LGD.  Two
scintillation counters are mounted on DEA, a window-frame counter
(CPVC) to distinguish between charged and neutral particles
hitting DEA, and a rectangular counter (CPVB) which covers the
hole in the DEA and is used, in conjunction with CPVC, to veto
charged tracks in the all-neutral trigger. Beyond the magnet, and
just upstream of the LGD, is a final drift chamber (TDX4)
consisting of two X-planes, and two scintillation counters (BV and
EV) for vetoing non-interacting beam particles and
elastic-scattering events. Further details regarding the equipment
are given elsewhere~\cite{a0paper}.

\section{Data Selection and Properties}

The trigger for the $\eta \pi^+ \pi^-$\ topology required a
\^{C}erenkov-tagged $\pi^-$ incident on the target, two charged
tracks emerging forward from the target, no charged recoil track,
and an effective mass greater than that of the $\pi^0$ in the LGD
as determined by a hardware processor. Forty eight million
triggers of this type were taken. From these, a final sample of
 events consistent with reaction \ref{reaction} was selected
by requiring:
\begin{itemize}
\item
less than 20 MeV in the CsI to enhance recoil neutron events over $N^*$
events;
\item
 exactly two photons ($\eta$) reconstructed in the LGD;
\item
a reconstructed beam track;
\item
two forward
charged tracks of opposite charge;
\item
no recoil charged track;
\item
 a 3-constraint SQUAW~\cite{squaw} kinematic fit to reaction \ref{reaction}
with a confidence level greater than $7\%$;
\item
 $|t| < 3$ \PGEV\ after kinematic fitting, where $t$ is defined as
the magnitude of the four-momentum transfer squared between the
target proton and the neutron in the final state.
\end{itemize}

\begin{figure}
\centerline{\epsfig{file=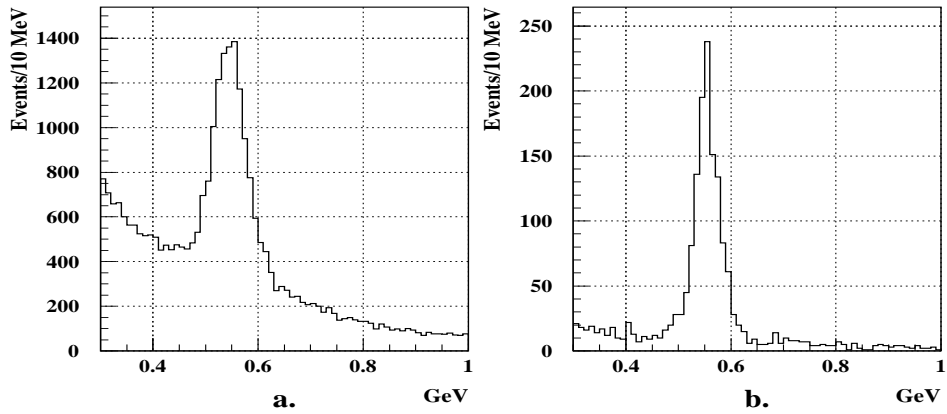,height=2.7in,width=5.5in,angle=0}}
    \caption{Two $\gamma$ invariant mass distribution
    (a) before cuts, (b)
    after data selection cuts.
\label{twogmass}}
\end{figure}

The two-gamma mass distributions for about $10\%$ of the data are
shown before and after the above data selection cuts in
Figs.~\ref{twogmass}(a) and (b) respectively.
The $\eta$ signal is nearly background free after cuts.

\begin{center}
\begin{figure}
    \psfig{figure=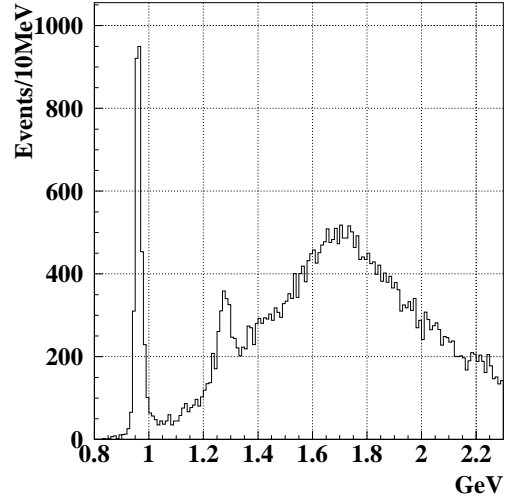,width=7.5cm,angle=0}
    \caption{$\eta \pi^+ \pi^-$\ three-body mass distribution
    (not corrected for acceptance).
\label{mass}}
\end{figure}
\end{center}

The $\eta\pi^+\pi^-$ mass distribution for these events is shown
in Figure~\ref{mass}. The $\eta^{'}(958)$ is evident.  When fit
with a Gaussian, a mass of 961~${\rm MeV}/c^2$ with $\sigma$
=10~${\rm MeV}/c^2$ is obtained. This provides a measure of the
mass resolution of the apparatus in the 1000~${\rm MeV}/c^2$ mass
region after kinematic fitting. An enhancement in the 1300 ~${\rm
MeV}/c^2$ region\footnote{A detailed description of the Dalitz
plot in this region is given is elsewhere~\cite{a0paper}.} is also
observed, which, when fit to a Gaussian plus a linear background,
yields a mass of 1278~${\rm MeV}/c^2$ and $\sigma$ = 20~${\rm
MeV}/c^2$.

\begin{figure}
\centerline{\epsfig{file=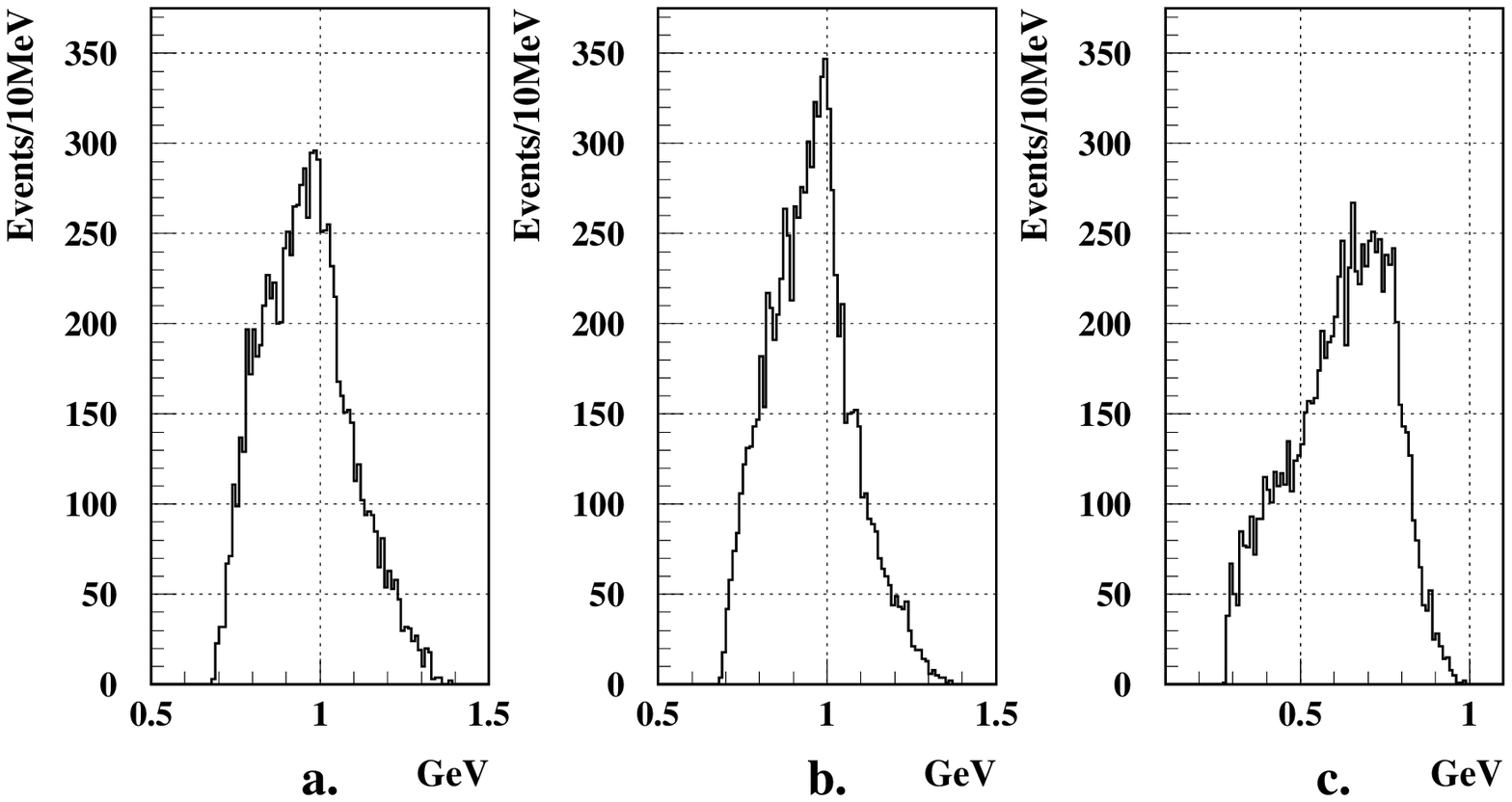,height=2.7in,width=5.5in,angle=0}}
    \caption{Two-body mass distributions: a) $\eta \pi^-$\,
     b) $\eta  \pi^+$\, and
     c) $\pi^+ \pi^-$\ for
     the $\eta \pi^+ \pi^-$\ mass region between 1200 and 1540~${\rm MeV}/c^2$.
\label{pmass}}
\end{figure}

In Figs.~\ref{pmass}a,~\ref{pmass}b and~\ref{pmass}c we show the
$\eta\pi^-$, $\eta\pi^+$ and $\pi^+\pi^-$ effective mass
distributions respectively for a three-body mass between 1200 and
1540~${\rm MeV}/c^2$. In Figs.~\ref{pmass}a and~\ref{pmass}b the
$a_0(980)$ peak is seen. The $\rho(770)$ peak in Fig.~\ref{pmass}c
is cut off on the high-mass side because of the limited phase
space available.

\begin{figure}
\centerline{\epsfig{file=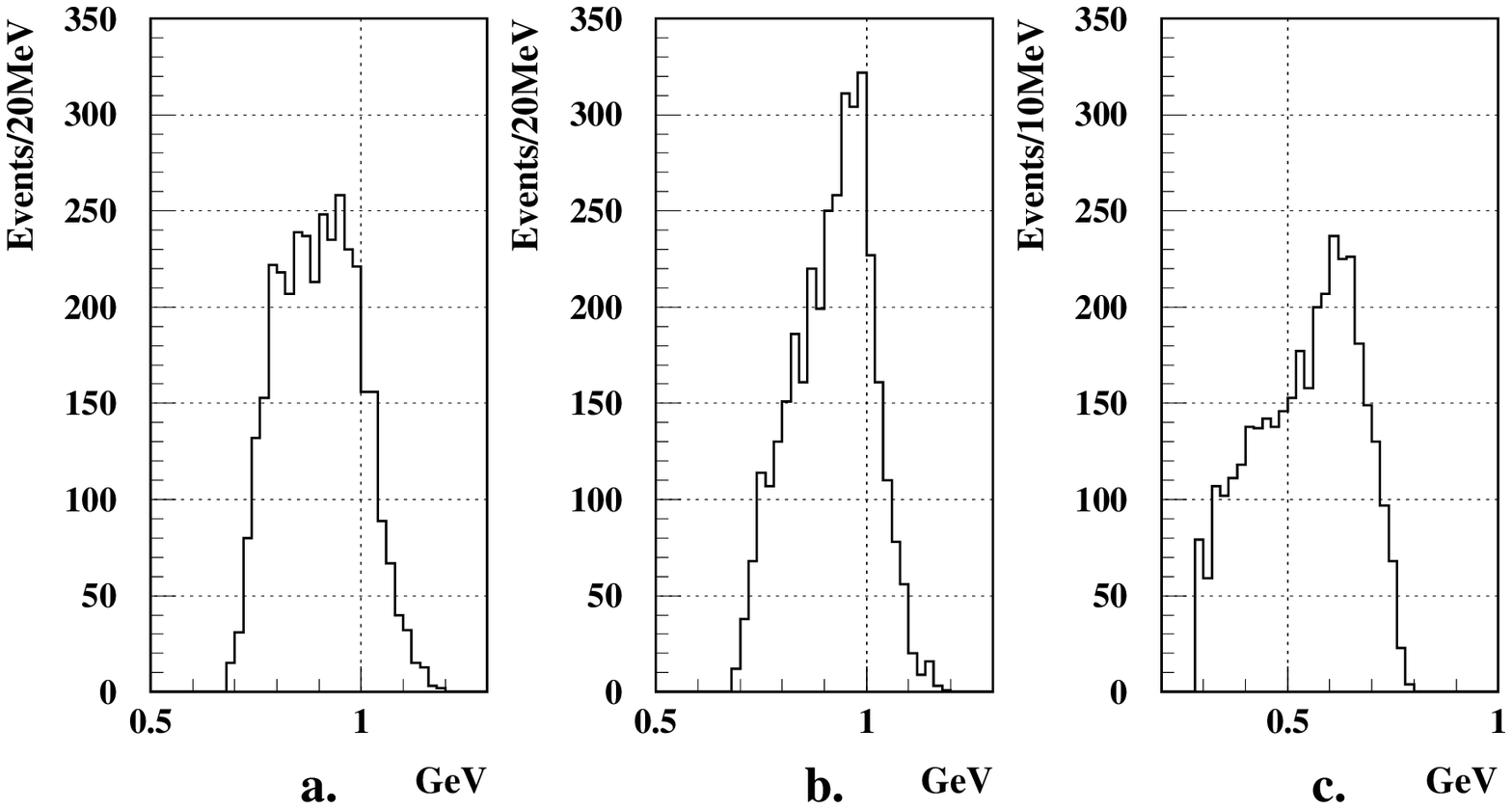,height=2.7in,width=5.5in,angle=0}}
    \caption{Two-body mass distributions: a) $\eta \pi^-$\,
     b) $\eta  \pi^+$\, and
     c) $\pi^+ \pi^-$\ for
     the $\eta \pi^+ \pi^-$\ mass region between 1200 and 1350~${\rm MeV}/c^2$.
\label{pmass_d}}
\end{figure}

The same distributions are shown in Fig.~\ref{pmass_d}(a)-(c) for
the low mass subset of the data between 1200 and 1350~${\rm
MeV}/c^2$. A very significant asymmetry between the $\eta\pi^-$
and the $\eta\pi^+$ distributions is evident. This asymmetry is
due to the interference between I=0 $a_0\pi$ and I=1 $\rho\eta$
states, and is well-described in the partial-wave analysis
described in the next section.

For the following analysis, 9,082 events were selected from the
above data set in the region $1205\le M(\eta\pi\pi)\le 1535~{\rm
MeV}/c^2$.

\section{Partial-wave analysis}

\subsection{Fitting Procedure}
\label{fitting}
 The formalism used in this analysis is based on
the papers of Chung~\cite{chung} and Chung and
Trueman~\cite{chungtru}. The analysis techique involved the use of
the reflectivity basis to describe the individual partial waves
and the maximization of an extended log likelihood function in the
fitting procedure. Fits are carried out independently in each
$\eta\pi^+\pi^-$ mass bin. The procedure and analysis programs are
described by Cummings and Weygand~\cite{newbnl}.

Due to the large number of possible partial waves accessible to
the $\eta \pi^+ \pi^-$ system, a complete analysis requiring all
possible isobars and partial waves is not practical given the
limited statistics. In principle one would like to include all
possible isobars: $\sigma, \rho, a_0, f_2, a_2$ and a large set of
partial waves ($J<4$). Because this analysis is limited to the
low-mass region, we can neglect the $a_2$\ and the $f_2$
isobars.\footnote{ The $f_2(1270)$ could in principle reach this
final state through the $a_2\pi$ mode, but this is highly
suppressed by phase space.} Furthermore, we choose to consider
only amplitudes with $J<2$ since there are no known states with
higher spin in this low-mass region which decay into $\eta\pi\pi$.

An incoherent isotropic background was included in some trial
fits, but it was not used in the final fit. This type of
background is, except for the $\pi\pi$ mass dependence of the
amplitude, similar to a non-interfering $J=0$, $\sigma \eta$\
partial wave, making them quite difficult to differentiate.

In order to determine whether both spin-flip and non-flip
amplitudes at the baryon vertex are required to fit the data, fits
were attempted for both rank 1 and 2. (If both types of amplitudes
are not necessary, the rank 1 fit will give a good description of
the data.) The likelihood function was improved greatly when the
fit rank was increased to two. In addition, rank 1 fits to the
data were found to become unstable in the absence of a background
wave. We conclude that a rank 2 fit is required to fit the data;
 rank 1 fits were not used.

The $\rho$ isobar was modeled by a relativistic Breit-Wigner
amplitude with parameters extracted from the Particle Data
Book~\cite{book:pdb}. For the final fit the $a_0$ isobar was
modeled as a Breit-Wigner form with a mass of 980~${\rm MeV}/c^2$
and a width of 72~${\rm MeV}/c^2$~\cite{a0paper}. The $\pi\pi$
S-wave (the $\sigma$) was represented by a parameterization of the
\pip \pim\ S-wave provided by K. L. Au {\em et al.}~\cite{th:AMP}.
Alternate parameterizations of the $a_0$\ and the $\sigma$\ were
explored~\cite{amsler}. However it was found that the particular
choice of parameterization had little effect on the final results.

To determine the appropriate waves for the 1200-1350~${\rm
MeV}/c^2$ region, a fit was performed using a coarse bin width of
50~${\rm MeV}/c^2$ with all waves with $J<2$ included. Waves were
then discarded from the fit if their removal had little effect on
the value of the likelihood function ($|\Delta {\cal L}| < 5$).
Selected waves were then re-introduced in the fit to insure the
stability of the solution. In total, several hundred different
sets of partial waves were fit. For each combination of partial
waves, the binning, $t$\ cuts, and starting values of the fit
parameters were varied to insure stability of the final solution.
The set of waves chosen for the final fit is shown in
Table~\ref{t:lowwaves}.

For the final fit, a  bin width of 30~${\rm MeV}/c^2$ was chosen.
This was a compromise between achieving adequate statistics in
each mass bin and acquiring the best possible resolution in the
entire mass region 1200-1540${\rm MeV}/c^2$. The starting values
of the fit were randomly chosen and the entire spectrum was re-fit
several times to insure stability with the finer bin width. For
bin widths smaller than 30~${\rm MeV}/c^2$ the fits often became
unstable, converging to different solutions depending upon the
starting values.

It is interesting to note that the final waves selected for the
low-mass region are consistent with those used by Stanton {\em et
al.}~\cite{ch1:stanton} and by Fukui {\em et
al.}~\cite{ch1:fukui}. The only exception is that we do not use
 a $1^{+-} \; a_0 \pi$ wave. While the inclusion
of this wave in the fit for the low-mass region is reasonable,
providing a natural explanation for the odd-even isospin
interference observed in the data, we found that the fit could not
distinguish this wave from the $1^{+-} \; \rho \eta$ wave in this
mass range. Due to the unambiguous presence of the $\rho \eta$\
partial wave at higher mass, it was decided not to include the
$1^{+-} \; a_0 \pi$\ partial wave in the final set.

\subsection{Results of Partial Wave Analysis}

\subsubsection{$J^{PC}=0^{-+}  $ Partial Waves}

\begin{figure}[b!] 
\centerline{\epsfig{file=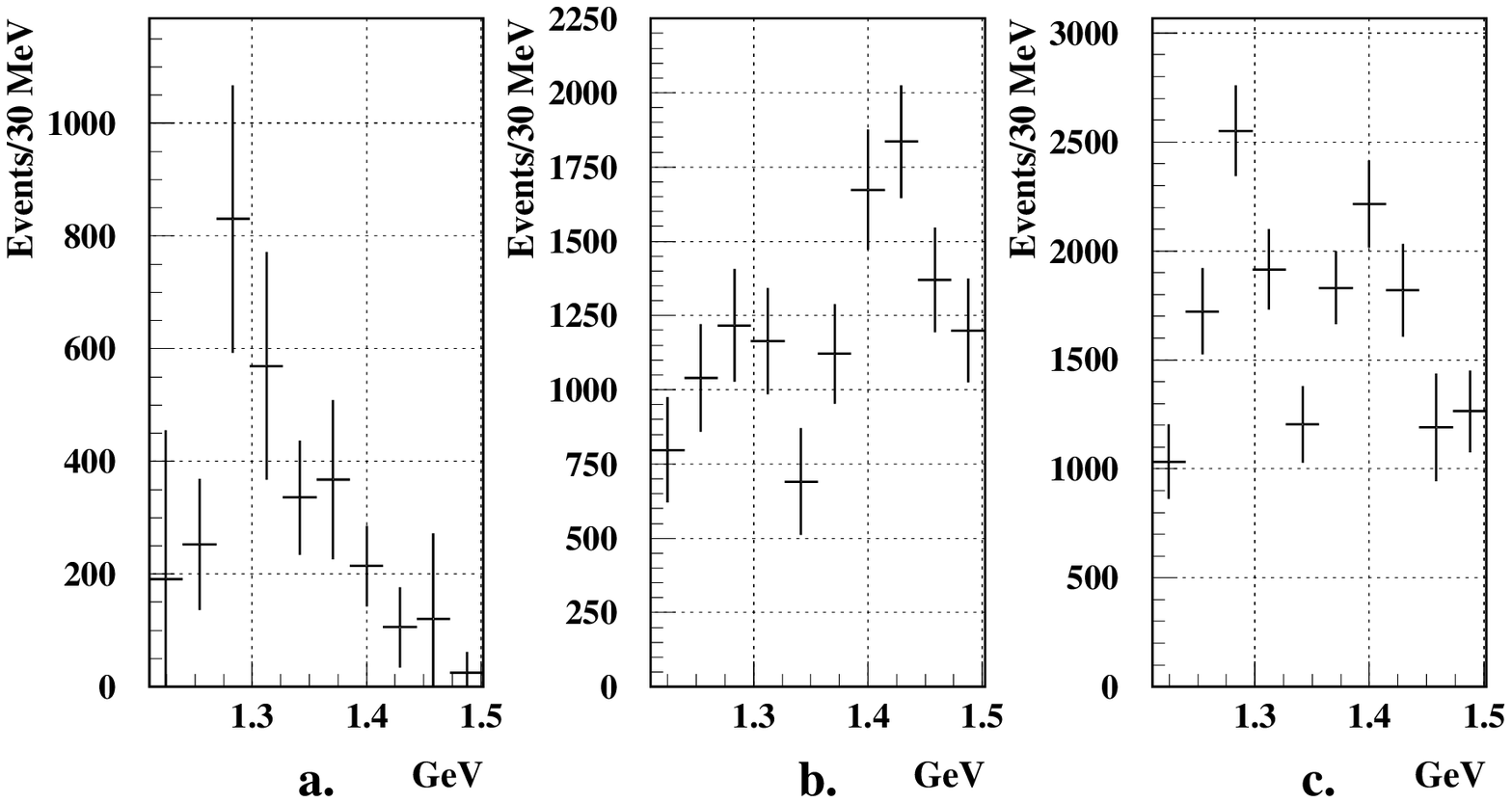,height=2.7in,width=5.5in}}
\vspace{10pt}
    \caption{ a: $0^{-+} a_0 \pi$ intensity, b: $0^{-+} \sigma \eta$ intensity,
c: Total $0^{-+}$ intensity}
\label{0-+etpipi}
\end{figure}

The fitted intensity distribution for the $0^{-+}~a_0\pi$ wave as
a function of $\eta\pi^+\pi^-$ effective mass is shown in
Fig.~\ref{0-+etpipi}a. A sharp peak at $\approx$ 1300~${\rm
MeV}/c^2$ is evident, consistent with the observation of
$\eta(1295) \rightarrow a_0 \pi$. Some intensity is seen extending
out to 1400~${\rm MeV}/c^2$. It should be noted that the majority
of the signal for this wave comes from the second rank of the fit.
This indicates a different production mechanism than that for the
$1^{+-}$\ and the $1^{++}$\ waves (which are produced dominantly
in the first rank) and means that these latter waves do not
interfere with the $0^{-+}$ wave (see discussion below).

 As shown in Fig.~\ref{0-+etpipi}b, the $0^{-+}\sigma \eta$ wave is
double-peaked, with structure  suggestive of $\eta(1295)$ and
$\eta(1440)$ production.  The dominant nature of the structure
seen in the high-mass region in this wave is somewhat inconsistent
with previous analyses~\cite{ch1:fukui,ch1:ando} which observe the
presence of a $\sigma \eta$ decay  of the $\eta(1440)$, but do not
see it as dominant.

A large fraction of the $0^{-+}$\ signal occurs in the second
rank, especially for the $\sigma \eta$ partial waves in the
high-mass region. Because these waves do not interfere with the
other dominant waves in the fit, reliable relative phase motion
could not be obtained.

The $a_0\pi$ and $\sigma \eta~0^{-+}$\ waves were added coherently
in each rank and then summed incoherently. The result is shown in
Fig.~\ref{0-+etpipi}c. The $\eta(1295)$ and $\eta(1440)$ peaks are
clearly visible. The spectrum was fit with two spin-0 Breit-Wigner
forms plus a quadratic background. Fitted values of the masses and
widths are given in Table~\ref{masstable}. In addition, the $a_0
\pi / \sigma \eta$ branching ratio were determined from
Fig.~\ref{0-+etpipi} for both the $\eta(1295)$ and the
$\eta(1440)$. These values are also given in
Table~\ref{masstable}.

\subsubsection{$J^{PC}=1^{++} $ Partial Waves}

\begin{figure}
\centerline{\epsfig{file=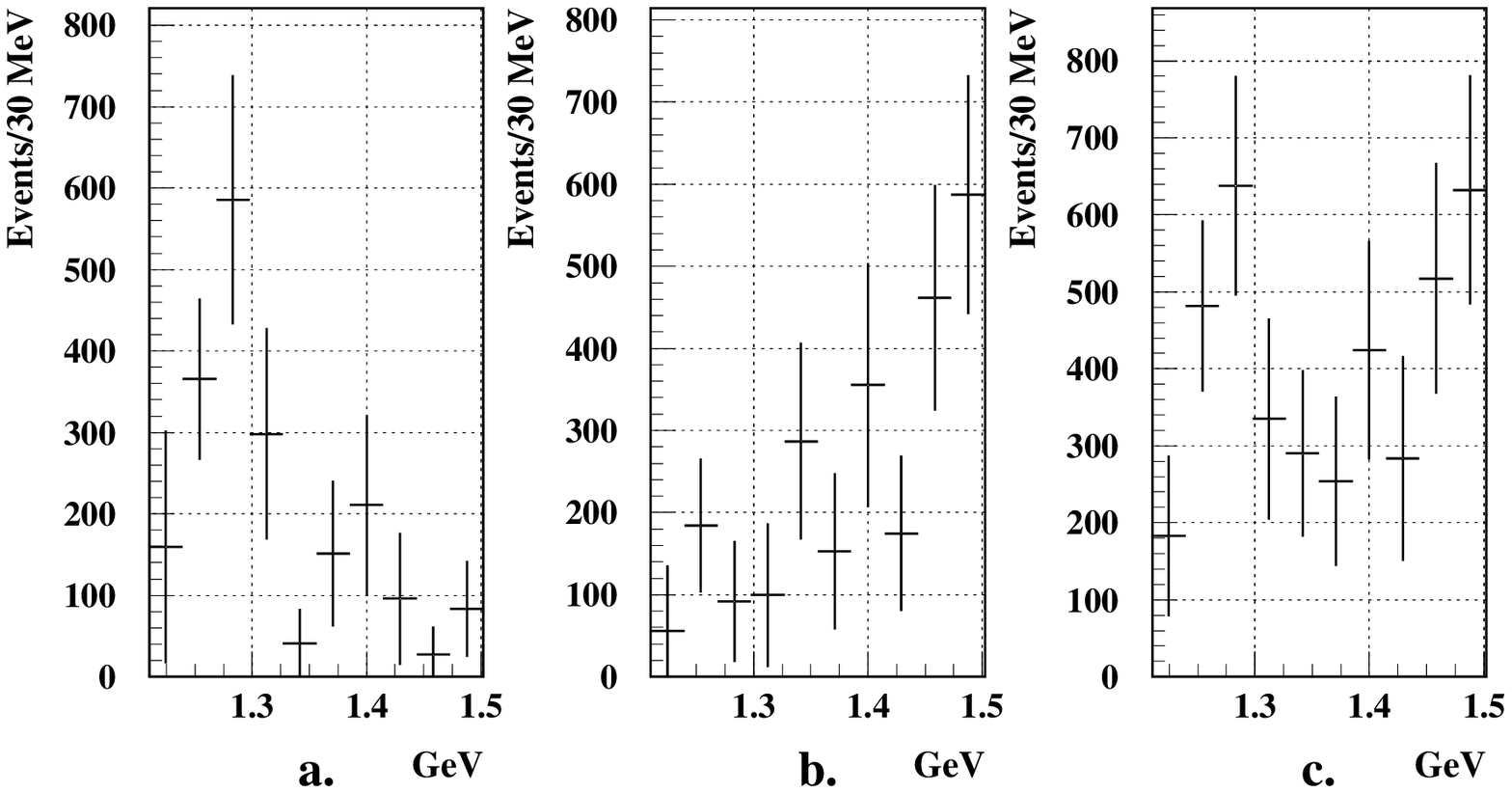,height=2.7in,width=5.5in}}
   \caption{a: $1^{++} a_0 \pi$ intensity,
b: $1^{++} \sigma \eta$ intensity,
c: Total $1^{++}$ intensity
\label{1++etpipi}}
\end{figure}

In Fig.~\ref{1++etpipi}a the $1^{++}\ a_0 \pi$ partial wave
intensity distribution is shown. This wave shows  evidence for the
$f_1(1285)$. The amount of $f_1$\ signal is comparable to  the
 $0^{-+}$\ signal in the $\ a_0 \pi$ channel. No significant structure is
observed at higher mass.

In Fig.~\ref{1++etpipi}b  the $1^{++}\ \sigma \eta$\ partial wave
intensity distribution is shown. This wave does not show any
structure, but was necessary to the fit for bins above 1450~${\rm
MeV}/c^2$.

In Fig.~\ref{1++etpipi}c  the coherent sum of the $1^{++}$\
partial waves is shown. This sum displays a peak in the vicinity
of the $f_1(1285)$\ and a rise at high mass.  A comparison of
Fig.~\ref{1++etpipi}c and Fig.~\ref{0-+etpipi}c reveals that the
majority of the signal strength in the 1200-1500~${\rm MeV}/c^2$
mass region arises not from $1^{++}$\ partial waves, but from the
$0^{-+}$ wave. This observation is especially important for the
low-mass region because several previous
analyses~\cite{ch1:corden,ch1:gurtu} for branching ratio estimates
assumed that the low-mass region was dominated by the $1^{++}$
wave.  We find that the ratio of the $1^{++}$ intensity to the sum
of the $1^{++}$ and $0^{-+}$ intensities in the region
1235-1325~${\rm MeV}/c^2$ is $0.19\pm 0.06$.

The observed $ f_1(1285) $ signal in Fig.~\ref{1++etpipi}c is in
the same mass region with a very similar width as  the $ \eta
(1295) $. To eliminate the possibility that this $ 1^{++} $ $
(f_1) $ signal is an artifact, due to ``leakage" from the larger $
0^{-+} $ $ (\eta) $ signal, a Monte-Carlo study was performed. The
measured amplitudes, with the contribution due to the $ 1^{++} $
wave removed, were used to generate Monte-Carlo events taking into
account the experimental resolution, acceptance and statistics.
These events were then analyzed in exactly the same manner as the
data. The resulting $ 1^{++} $ intensity in the low-mass region
was found to be less than 2\% of the total signal, and consistent
with zero.  This leads to the conclusion that the observed signals
are not artifacts of the analysis or apparatus, and are, in fact,
two distinct resonances.

\subsubsection{$ \rho \eta$ Partial Waves}

\begin{figure}
\centerline{\epsfig{file=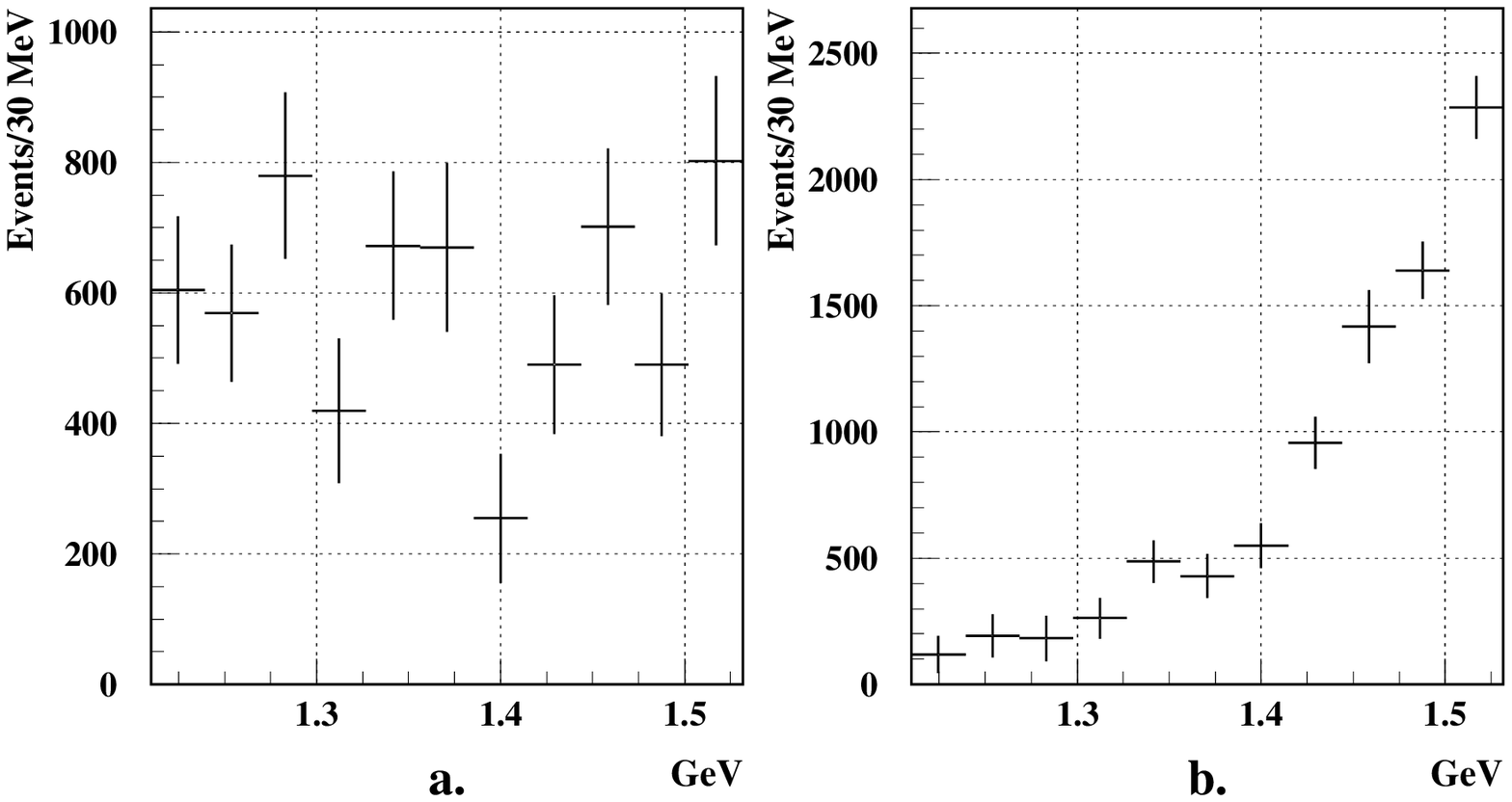,height=2.7in,width=5.5in}}
\caption{ a: $1^{+-} \rho \eta$ intensity, b: $1^{--} \rho \eta$ intensity (negative reflectivity partial wave)
\label{1xxetpipi}}
\end{figure}

 The intensity distribution for the
$1^{+-}\rho \eta$\ wave is shown in Fig.~\ref{1xxetpipi}a. This
wave was seen in all previous analyses and is significant in the
low-mass region. Previous experiments~\cite{ch1:fukui} have
claimed this wave to show evidence for production of the
$b_1(1235)$\ with a $\rho \eta$\ decay mode. However, we do not
observe any structure in the $1^{+-}$ wave to support this
conjecture. As mentioned earlier, this wave is, nevertheless,
essential for producing the $a_0^+ / a_0^-$\ asymmetry observed in
the data.

In Fig.~\ref{1xxetpipi}b  the $1^{--} \rho \eta$\ intensity
distribution is shown. This is the only negative-reflectivity
partial wave in our analysis, and it does not interfere with any
other partial waves in the fit. The wave steadily increases
throughout the low-mass region, consistent with its being the
low-mass tail of the $\rho(1700)$.

\section{Discussion of the $\lowercase{f}_1(1285)$ Branching Fractions}
\label{bfsection}

Production of the $\eta(1295)$ dominates the low-mass peak,
accounting for roughly 80\% of the signal. This observation has
implications on the $f_1 \rightarrow \eta \pi \pi$ branching
fraction. Previous
experiments~\cite{ch1:campbell,ch1:corden,ch1:defoix,ch1:gurtu}
have determined the $f_1 \rightarrow \eta \pi \pi$\ branching
fraction without the aid of a partial wave analysis under the
assumption that the low-mass peak consists of a single $f_1$ state
resting on top of an incoherent background. This assumption is
clearly incorrect, and values for the earlier determinations of
the branching fractions need to be corrected.

Corden {\em et al.}~\cite{ch1:corden} studied the reactions
\reaction\ and $\pi^- p \rightarrow K \overline{K} \pi n$ at
15~\EGEV. In their analysis they obtained a $K\overline{K} \pi /
\eta \pi \pi$ branching ratio in the low-mass region of
$0.5\pm0.2$\ without the aid of a partial wave analysis. It is
reasonable to assume that the relative production of $f_1(1285)$
and $\eta(1295)$ is the same in the present experiment as in that
of Ref.~\cite{ch1:corden} since the experiments are close in
energy and study the same final state.  It is also reasonable to
assume that the $K \overline{K} \pi$\ decay at low mass is
entirely due to $f_1(1285)$ decay since this conclusion was
reached by a partial wave analysis~\cite{ch1:birman} of the data.
Thus the $K\overline{K} \pi / \eta \pi \pi$\ branching ratio of
the $f_1(1285)$, as quoted by Corden {\em et al.}, should be
corrected by dividing it by the fraction of the low-mass peak
which is due to $f_1(1285)$ decay.  We thus obtain
$(0.5\pm0.2)/(0.19\pm0.06) = 2.6 \pm 1.4$ for this branching
ratio.

We can perform the same type of estimate  using, instead of our
own analysis,  the results of KEK-E179~\cite{ch1:fukui} for the
reaction \reaction\ at 8.95~\EGEV. In that experiment, the
fraction of the low-mass peak which is due to $f_1(1285)$ decay is
claimed to be 50\%. Again, using the results of Corden {\em et
al.} (although in this case the difference in the energies of the
experiments is larger), we obtain an alternate estimate of the $
K\overline{K} \pi / \eta \pi \pi$ branching ratio for $f_1$ decay
to be $1.0 \pm 0.4$.

We can estimate the effect which these results can have on the
$f_1(1285)$\ branching fractions by assuming that the low-mass
signal observed in the $K \overline{K} \pi$, $\gamma \rho^0$, and
$4 \pi$ decay modes is due only to $f_1(1285)$ decay. This is the
most reasonable for the $K \overline{K} \pi$\ mode as mentioned
above because the other two decay modes ($\gamma \rho^0$, $4 \pi$)
have not been as thoroughly investigated.\footnote{Of these modes
the $4 \pi$\ branching fraction is most suspect due to the large
number of interfering partial waves which contribute to a $4 \pi$
data set.} Nevertheless, using the PDG98 branching
ratios~\cite{book:pdb} of $0.271 \pm 0.016$ for $K \overline{K}
\pi / 4 \pi$ and $0.45 \pm 0.18$ for $\gamma \rho^0 / 2 \pi^+ 2
\pi^-$, the $f_1(1285)$ branching fractions can be calculated. (We
also assume the branching ratio for $4\pi/2 \pi^+ 2 \pi^-=3$ as
in~\cite{book:pdb}.) In Table~\ref{t:f1bf} we list the
$f_1(1285)$\ branching fractions derived by the above procedure.

Assigning systematic errors to these $f_1(1285)$ branching
fractions is difficult because of the undetermined uncertainties
in branching ratios for the $4 \pi$ and $\gamma \rho^0$\ decay
modes. However, it is clear from the above exercise that the
results from the present experiment and the KEK experiment for the
$f_1(1285)$\ branching fractions are consistent, and those listed
in the particle data book~\cite{book:pdb} need to be corrected.
The most significant result is the large reduction in the
$f_1\rightarrow\eta \pi \pi$ branching fraction.

\section{Summary and Conclusions}

A partial wave analysis was performed on 9082 $\eta\pi^+\pi^-n$
events in the $1205\leq M(\eta\pi^+\pi^-)\leq 1535~{\rm MeV}/c^2$
mass interval. The analysis used a rank 2 fit with 30 ${\rm
MeV}/c^2$ bins and a set of 6 partial waves. The partial waves
used in the fit were: $0^{-+} a_0 \pi$, $0^{-+} \sigma \eta$,
$1^{+-} \rho \eta$, $1^{++} a_0 \pi$, $1^{++} \sigma \eta$\ and
$1^{--} \rho \eta$.

The low-mass region was found to include a large contribution from
the $0^{-+}$\ wave which indicates the production of $\eta(1295)$.
Evidence of $f_1(1285)$ production was seen in the $1^{++}$ wave.
The fact that the region is dominated by $\eta(1295)$ production
leads to significant changes in the $f_1(1285)$ branching
fractions as discussed in Section~\ref{bfsection}.

The $\eta(1295)$ was seen to decay to both $a_0 \pi$ and $\sigma
\eta$. The $a_0 \pi / \sigma \eta$\ branching ratio for
$\eta(1295)$\ was estimated to be $0.48 \pm 0.22$. The mass and
width of the $\eta(1295)$ were determined to be $1282 \pm 5~{\rm
MeV}/c^2$ and $66 \pm 13~{\rm MeV}/c^2$ respectively. This result
is consistent with the PDG98 summary of $\eta(1295)$ mass and
width of $1297 \pm 2.8~{\rm MeV}/c^2$ and $53 \pm 6~{\rm
MeV}/c^2$, respectively.

The high-mass region is dominated by a large $0^{-+}  \sigma \eta$
signal present in the second rank of the fit. This signal is
consistent with production of a single state, the $\eta(1440)$.
The mass and width of the $\eta(1440)$\ are estimated to be $1404
\pm 6~{\rm MeV}/c^2$ and $80 \pm 21~{\rm MeV}/c^2$. This result is
consistent with the PDG98~\cite{book:pdb} weighted average value
for the mass and width of the $\eta(1440)$ determined from the
$\eta\pi\pi$ mode of $1405 \pm 5~{\rm MeV}/c^2$ and $56 \pm 7~{\rm
MeV}/c^2$, respectively.

The $\eta(1440)$ has been previously observed in the reaction
$\pi^- p \rightarrow K \overline{K} \pi n$, in $p\overline{p}$
annihilation, and in the radiative decay of $J/\psi$, with decays
in the $a_0 \pi$ and $K\overline{K^{\ast}}$  modes. Studies of the
\reaction\ reaction have yielded both a $\sigma \eta$\ and an $a_0
\pi$ component of the $\eta(1440)$. In the present analysis, it is
found that the $\sigma \eta$ decay dominates, while the KEK
analyses~\cite{ch1:fukui,ch1:ando}
 suggest a larger $a_0 \pi$ component. The estimate
of the  $a_0 \pi / \sigma \eta$ branching ratio for $\eta(1440)$
from the present analysis is $0.15 \pm 0.04$. The systematic
errors are unassigned, but assumed to be large due to the
difficulty of the fit in distinguishing $0^{-+} a_0 \pi$ and
$0^{-+} \sigma \eta$ waves from each other.

In addition to the $f_1(1285)$, $\eta(1295)$\ and the $\eta(1440)$\
contributions, a large,
relatively structureless signal in the
$1^{+-}\rho \eta$ wave was observed throughout
the low mass region. This wave has also been observed in all
previous partial wave analyses of the
\reaction\ system. There is no obvious
resonance interpretation of this structure, but its
 presence is
required to account for the large $a^+_0 / a^-_0$\ production
asymmetry seen in the low mass region. A $1^{--} \rho \eta$
partial wave, consistent with the low-mass tail of the
$\rho(1700)$, is also seen.

We would like to express our deep appreciation to the members of the MPS group.
Without their outstanding efforts, the results presented here could not have
been obtained.  We would also like to acknowledge the invaluable assistance of the
staffs of the AGS and BNL, and of the various collaborating institutions.
This research was supported in part by the National Science Foundation and
the US Department of Energy.

\begin{table}
\caption{Partial Waves Used in  Final Fit \label{t:lowwaves}}
\begin{center}
\begin{center}
\begin{tabular}{|c|c|c|c|c|c|}
$Isospin$ & $J^{PC}$ & Isobar & $l$ & $m$ & $\epsilon$ \\ \hline
\hline 1   & $1^{--}$ & $\rho$ & 1   & 0   & $-1$ \\ \hline 0   &
$0^{-+}$ & $a_0$  & 0   & 0   & $+1$ \\ \hline 0   & $0^{-+}$ &
$\sigma$  & 0   & 0   & $+1$ \\ \hline 0   & $1^{++}$ & $a_0$  & 1
& 0   & $+1$ \\ \hline 0   & $1^{++}$ & $\sigma$  & 1   & 0   &
$+1$ \\ \hline 1   & $1^{+-}$ & $\rho$ & 0   & 0   & $+1$ \\
\end{tabular}
\end{center}
\end{center}
\end{table}

\begin{table}
\caption{ Properties of the $J^{PC} = 0^{-+}$  States
\label{masstable}}
\begin{center}
\begin{center}
\begin{tabular}{|c|c|c|c|}
      & Mass (\MGEV) & Width (\MGEV)& $a_0\pi/\sigma\eta$ Branching Ratio \\ \hline
$\eta(1295)$ & $1.282 \pm 0.005$& $0.066 \pm 0.013 $ & $0.48 \pm
0.22$   \\ \hline $\eta(1440)$ & $1.404 \pm 0.006$& $0.080 \pm
0.021 $ & $0.15 \pm 0.04$   \\
\end{tabular}
\end{center}
\end{center}
\end{table}

\begin{table}
\caption{ $f_1$ Branching Fractions \label{t:f1bf}}
\begin{center}
\begin{center}
\begin{tabular}{|c|c|c|c|}
Decay Mode      & PDG~\cite{book:pdb} & BNL-E852 &
KEK-E179~\cite{ch1:fukui,ch1:ando} \\ \hline \hline $4\pi$ & $35
\pm 4$\%         & $65 \pm 4 $\% & $59 \pm 5 $\%   \\ \hline $\eta
\pi \pi$  & $50 \pm 18$\%        & $7 \pm 3  $\%  & $16 \pm 5 $\%
\\ \hline $\gamma \rho^0$ & $5.4 \pm 1.2$\%      & $10 \pm 4 $\% &
$9 \pm 3 $ \%   \\ \hline $K \overline{K} \pi$ & $9.6 \pm 1.2$\% &
$18 \pm 1 $\%  & $16 \pm 1 $\% \\
\end{tabular}
\end{center}
\end{center}
\end{table}


\begin{references}
\bibitem{ch1:campbell} {J. H. Campbell {\em et al.},
Phys. Rev. Lett. {\bf22}, 1204 (1969)}.
\bibitem{ch1:corden} {M. J. Corden {\em et al.},
Nucl. Phys.  {\bf B144}, 153 (1978)}.
\bibitem{ch1:Dahl} {O. I. Dahl {\em et al.},
Phys. Rev. {\bf 164}, 1377 (1967)}.
\bibitem{ch1:nacasch} {R. Nacasch {\em et al.},
Nucl. Phys.  {\bf B203}, 203 (1978).}
\bibitem{ch1:defoix} {C. Defoix {\em et al.},
Nucl. Phys.  {\bf B44}, 125 (1972).}
\bibitem{ch1:gurtu} {A. Gurtu {\em et al.},
Nucl. Phys.  {\bf B151}, 181 (1979).}
\bibitem{book:pdb} {Particle Data Group: C. Caso {\em et al.},
The European Physical Journal {\bf C3}, 1 (1998).}
\bibitem{ch1:stanton} {N. R. Stanton {\em et al.}, Phys. Rev. Lett. {\bf42}, 346
(1979)}.
\bibitem{ch1:fukui} {S. Fukui {\em et al.}, Phys. Rev. B {\bf267}, 293
(1991)}.
\bibitem{ch1:ando} {A. Ando {\em et al.}, Phys. Rev. Lett. {\bf57}, 1296
(1986)}.
\bibitem{crittendon} {B. Brabson {\em et al.}, Nucl. Instr. \& Meth A {\bf332}, 419
(1993)}.
\bibitem{crittendon2} {R. R. Crittenden, Nucl. Instr. \& Meth A, {\bf387}, 377
(1997)}.
\bibitem{csi} {T. Adams {\em et al.}, Nucl. Instr. \& Meth A {\bf386}, 617
(1996)}.
\bibitem{tcyl} {Z. Bar-Yam {\em et al.}, Nucl. Instr. \& Meth A {\bf342}, 398
(1994)}.
\bibitem{a0paper} {S. Teige {\em et al.}, Phys. Rev. {\bf D59}, 012001
(1999)}.
\bibitem{squaw} {O. I. Dahl {\em et al.}, ``SQUAW kinematic fitting program'',  Univ. of California, Berkeley Group A programming note P-126, unpublished (1968).}
\bibitem{chung}  {S.U. Chung, ``Formulas for Partial-Wave Analysis'',
Brookhaven BNL-QGS-93-05, unpublished (1993).}
\bibitem{chungtru}{S.U. Chung and T.L. Trueman, Phys. Rev. D {\bf111}, 633
(1975)}.
\bibitem{newbnl} {J. P. Cummings and D. P. Weygand, ``The New BNL Partial Wave Analysis Programs'', BNL-64637, unpublished
(1997).}
\bibitem{th:AMP} {K. L. Au {\em et al.}, Phys. Rev. D {\bf35}, 1633
(1986)}.
\bibitem{amsler} {C. Amsler {\em et al.}, Phys. Lett. B {\bf335}, 425
(1995)}.
\bibitem{ch1:birman} {A. Birman {\em et al.}, Phys. Rev. Lett. {\bf61}, 1557
(1988)}.
\end{references}
\end{document}